\documentclass{aa}  
\usepackage{comment}
\usepackage{graphicx}
\usepackage{txfonts}
\usepackage[para,online,flushleft]{threeparttable}
\usepackage{array}
\usepackage{booktabs}
\newcommand{\diff}{\ensuremath{\mathrm{d}}}
\begin{document} 

   \title{On the possible jet contribution to the $\gamma$-ray luminosity in NGC\,1068}

   \author{S.~Salvatore
          \inst{1, 2}%\inst{2}
          \and
          B.~Eichmann \inst{1, 2}%\inst{2} 
          \and X.~Rodrigues \inst{2, 3, 4}%\inst{3}\inst{4} 
          \and R.-J.~Dettmar \inst{2,4}%\inst{4} 
          \and J.~Becker Tjus \inst{1, 2, 5}
          %\inst{2}\inst{5}
          }

   \institute{Ruhr-Universit\"at Bochum, Fakult\"at f\"ur Physik und Astronomie, Theoretische Physik IV, 44780 Bochum, Germany   \and Ruhr Astroparticle and Plasma Physics Center (RAPP Center), 44780 Bochum, Germany \and European Southern Observatory, Karl-Schwarzschild-Straße 2, 85748 Garching bei München, Germany \and Ruhr-Universit\"at Bochum, Fakult\"at f\"ur Physik und Astronomie, Astronomisches Institut (AIRUB), 44780 Bochum, Germany \and Department of Space Earth and Environment, Chalmers University of Technology, 412 96 Gothenburg, Sweden}

\abstract
{NGC\,1068 is a nearby widely studied Seyfert II galaxy presenting radio, infrared, X- and $\gamma$-ray emission as well as strong evidence for high-energy neutrino emission. Recently, the evidence for neutrino emission could be explained in a multimessenger model in which the neutrinos originate from the corona of the active galactic nucleus (AGN). In this environment $\gamma$-rays are strongly absorbed, so that an additional contribution from e.g.\ the circumnuclear starburst ring is necessary. In this work, we discuss whether the radio jet can be an alternative source of the $\gamma$-rays between about $0.1$ and $100$ GeV as observed by Fermi-LAT. In particular, we include both leptonic and hadronic processes, i.e.\ accounting for inverse Compton emission and signatures from $pp$ as well as $p\gamma$ interactions. In order to constrain our calculations, we use VLBA and ALMA observations of the radio knot structures, which are spatially resolved at different distances from the supermassive black hole. 
%%%Previously:
%Our results show that only the radio knot closest to the central engine can, under certain conditions, explain the whole Fermi-LAT signal. For that, a rather weak magnetic field is necessary, of strength $\lesssim 0.3\,\text{mG}$, as well as a spectral softening of the relativistic electron distribution at about $10\,\text{GeV}$. We discuss a possible scenario for this emission site, but apart from that we conclude that this outflow alone cannot explain the observed $\gamma$-ray flux, neither closer to the foot of the jet nor overall. 
%%%NEW:
Our results show that the best leptonic scenario for the prediction of the Fermi-LAT data is provided by the radio knot closest to the central engine. For that a magnetic field strength $\sim 1\,\text{mG}$ is needed as well as a strong spectral softening of the relativistic electron distribution at $(1-10)\,\text{GeV}$. 
%However, we show that such a strong spectral break cannot be achieved through the main interaction processes happening in such knot (synchrotron, IC, bremsstrahlung). 
However, we show that neither such a weak magnetic field strength nor such a strong softening is expected for that knot. A possible explanation for the $\sim$ 10 GeV $\gamma$ rays can be provided by hadronic pion production in case of a gas density $\gtrsim 10^4\,\text{cm}^{-3}$. Nonetheless, this process cannot contribute significantly to the low energy end of the Fermi-LAT range. We conclude that the emission sites in the jet are not able to explain the $\gamma$-rays in the whole Fermi-LAT energy band. 

%We conclude that the production of $\gamma$-rays in the whole Fermi-LAT energy range from these radio knots is quite unlikely.
%\textbf{Alternative version with a stronger/clearer statement:}
%But at $\sim$ 10 GeV a $\gamma$-ray contribution by hadronic pion production is possible if we account for a dense gas target in the emission sites. Nonetheless, the hadronic scenario does not contribute to the low energy end of the Fermi-LAT range. 

%results naturally for this emission site and present a best-fit scenario that properly explains the observed radio and $\gamma$-ray data. 
}


   \maketitle

\section{Introduction}

The nearby starburst-Seyfert II galaxy NGC\,1068, located at a distance of $D_\text{L}$ = 10.1 $\pm$ 1.8 Mpc \citep{tully2009extragalactic}, is not only one of the first and brightest galaxies of its kind studied by \cite{Seyfert1943}, but also one of the first sources to show strong evidence for neutrino emission, as observed by the IceCube experiment \citep{icecube2022evidence}. NGC\,1068 shows also clear non-thermal emission in the $\gamma$-ray band up to some tens of GeV \citep{a5:fermi_starbursts2012}. The observation of a neutrino flux that is much higher than the gamma-ray flux resulted in the expectation that neutrinos must be produced in a $\gamma$-ray absorbed environment \citep{Inoue2020, a5:Murase+2020, Kheirandish+2021, eichmann2022solving, InoueCerrutiMuraseLiu2022}. Since the AGN corona is optically thick to $\gamma$-rays, it seems to be the most likely origin of the TeV neutrinos. The origin of the GeV $\gamma$-rays is less clear, since it may either be located in the starburst ring \citep{eichmann2022solving}, the torus \citep{InoueCerrutiMuraseLiu2022}, or the radio jet \citep{Lenain+2010}. In this work, we will be testing the hypothesis that the jet of NGC\,1068 contributes to the GeV $\gamma$-ray emission, based on radio observations of the jet and a physical model of the multi-wavelength non-thermal emission.

Radio observations \citep{Gallimore+2004,Michiyama_2022} show distinct emission sites at some tens of parsec as well as a few hundreds of parsec from the supermassive black hole of NGC\,1068. These are commonly associated with its jet that is inclined to the line of sight by about 45$^{\circ}$. Hereof, the most distant emission site shows spatial widening, including four bright knot structures according to ~\cite{Michiyama_2022}, and the typical lobe structure that marks the head of the jet. Here the whole structure seems to have hit the interstellar medium (ISM) and therefore presents an overall shock feature. Moreover, all of the observed radio structures can be inferred by spherical symmetric knot structures with a radius $r_{\rm k}$ on the order of about a few parsecs. According to \cite{roy2000slow} these knots are streaming non-relativistically away from the central engine with at most $v \sim 0.075c$.

Without accounting for the spatial features of the radio emission sites, \cite{Lenain+2010} has provided a phenomenological explanation of the observed $\gamma$-ray emission at GeV energies. They adopt an emission site at a distance from the torus of $65\,\text{pc}$, with a radius of $r_\text{k} = 7\,\text{pc}$ and a magnetic field of $0.1\,\text{mG}$ that however, is not observed in the radio band. Using relativistic electrons with a broken power-law energy distribution they manage to explain observed GeV $\gamma$-rays by inverse Compton (IC) scattering of IR photons from the torus. Based on the given radio power, the expected jet power $P_\text{jet}$ according to  \cite{cavagnolo2010relationship} is at the order of at most $10^{43}\,\text{erg}\,\text{s}^{-1}$. Thus, the jet of NGC\,1068 is comparably weak to produce the observed $\gamma$-ray luminosity between about $0.1$ and $100\,\text{GeV}$ of about $10^{41}\,\text{erg}\,\text{s}^{-1}$. 

In this work, we will account for the observational details of the individual radio knots and investigate if leptonic or hadronic emission in these knots can actually explain the observed $\gamma$-ray signal. 
The paper is structured as follows: 
in Sect.~\ref{sec:2} we introduce the dominant emission scenarios for the previously introduced emission sites, and in Sect.~\ref{sec:results} we apply them to the case of NGC\,1068. In Sect.~\ref{sec:disc&conc} we conclude and discuss whether or not the observed radio knots can also be considered as origin of the observed $\gamma$-ray signal.

%%AAAAAAAAAAAAA

\section{Theoretical toolbox}\label{sec:2}
%We limit our analysis to the following scenarios: (i) a leptonic scenario, where relativistic electrons inverse Compton (IC) scatter low-energy photons towards higher energies, and (ii) a hadronic scenario, where relativistic protons interact inelastically with these target photons via photomeson production as well as (iii) hadronic pion production where these protons interact with a background gas.

\subsection{Leptonic model}\label{lept}

In the presence of a single relativistic electron (with $\beta\simeq 1$), its emitted synchrotron power is given by $P_\text{syn} = 4\sigma_\text{T}c\gamma_\text{e}^2U_\text{B}/3$, where $\sigma_T$ denotes the Thomson cross section and $U_\text{B}=B^2/8\pi$ yields the magnetic energy density. 
Moreover, the spectral energy distribution of the emitted power shows a maximum at a frequency $\nu_{\rm syn}\simeq \gamma_{\rm e}^2\,\nu_{\rm L}$, with the characteristic Larmor frequency $\nu_{\rm L}=eB/2\pi m_\text{e}c$. 
In a similar manner also the total emitted power by IC scattering of a monochromatic photon target distribution with an energy density $U_{\rm ph}$ can be given in the Thomson limit by $P_{\rm IC} = 4\sigma_\text{T}c\gamma_{\rm e}^2\,U_{\rm ph}/3$.  
In addition, it is well known in the Thomson regime that an initial photon with a frequency $\nu_0$ obtains a mean frequency of $\nu_{\rm IC}=4\gamma_{\rm e}^2 \nu_0/3$.

In case of a differential energy distribution of electrons according to $n_{\rm e}(\gamma_{\rm e})\propto \gamma_{\rm e}^{-q_{\rm e}}$, the emissivity $\epsilon(\nu)$ at a given frequency $\nu$, within an interval $\diff\nu$, results for both emission processes from those electrons with an appropriate energy $\gamma_{\rm e}$, within the interval $\diff\gamma_{\rm e}$, so that 
\begin{equation}
    \epsilon_{\rm syn}(\nu_{\rm syn})\,\diff\nu_{\rm syn} \simeq \frac{1}{4\pi}\,P_{\rm syn}\left(\gamma_{\rm e}=\sqrt{\frac{\nu_{\rm syn}}{\nu_{\rm L}}}\right)\,n_{\rm e}\left(\gamma_{\rm e}=\sqrt{\frac{\nu_{\rm syn}}{\nu_{\rm L}}}\right)\,\diff\gamma_{\rm e}\,,
    \label{eq:syn_emis}
    \end{equation}
    and
    \begin{equation}
    \epsilon_{\rm IC}(\nu_{\rm IC})\,\diff\nu_{\rm IC} \simeq \frac{1}{4\pi}\,P_{\rm IC}\left(\gamma_{\rm e}=\sqrt{\frac{3\nu_{\rm IC}}{4\nu_{0}}}\right)\,n_{\rm e}\left(\gamma_{\rm e}=\sqrt{\frac{3\nu_{\rm IC}}{4\nu_{0}}}\right)\,\diff\gamma_{\rm e}\,.
\end{equation}

Hereby, we suppose that all the power is emitted at the frequencies $\nu_{\rm syn}$ and $\nu_{\rm IC}$, respectively, which is an appropriate estimate since the synchrotron and IC emissions show a peak at these characteristic frequencies. Note that these relations imply that all power is emitted at a typical frequency. 
With 
\begin{equation}
    \frac{\diff\gamma_{\rm e}}{\diff \nu_{\rm syn}}=\frac{\nu_{\rm syn}^{-1/2}}{2\nu_{\rm L}^{1/2}}\,,\quad\text{and}\quad \frac{\diff\gamma_{\rm e}}{\diff \nu_{\rm IC}}=\frac{\nu_{\rm IC}^{-1/2}}{2}\left( \frac{3}{4\nu_0}\right)^{1/2}\,
\end{equation}
the ratio of IC over synchrotron luminosity yields
\begin{equation}
    \frac{\nu_{\rm IC}L_{ \nu_{\rm IC}}}{\nu_{\rm syn}L_{\nu_{\rm syn}}}\equiv\frac{\nu_{\rm IC}\,\epsilon_{\rm IC}(\nu_{\rm IC})}{\nu_{\rm syn}\,\epsilon_{\rm syn}(\nu_{\rm syn})} = \left[ \frac{3}{4}\,\frac{\nu_{\rm IC}/\nu_{0}}{\nu_{\rm syn}/\nu_{\rm L}} \right]^{\frac{3-q_{\rm e}}{2}}\,\frac{U_{\rm ph}}{U_{\rm B}},
    \label{eq:LumRatio0}
\end{equation}
where the equality only holds in the Thomson limit, where $\nu_0\nu_{\rm IC}\ll (m_{\rm e}c^2)^2/h^2$. Hence, for a given target photon field (that is idealized to be monochromatic) and a known synchrotron flux spectrum at $\nu_{\rm syn}$, the resulting $\gamma$-ray luminosity $\nu_{\rm IC}L_{ \nu_{\rm IC}}$ from IC scattering depends on the magnetic field strength according to $\nu_{\rm IC}L_{ \nu_{\rm IC}} \propto B^{-(1+q_{\rm e})/2}$. 

Supposing that the energy within the emission site is at equipartition, an estimate of the corresponding magnetic field strength can be derived---similar to the the procedure in \cite{pacholczyk1970non}---as follows: the total electron energy budget of those electrons that predominantly contribute to the synchrotron luminosity $L_{\nu_{\rm syn}}$, i.e. those with $\gamma_\text{e} = \sqrt{\nu_{\rm syn}/\nu_{\rm L}}$, can be estimated by
\begin{equation}\label{Ee1}
    E_\text{e} \simeq m_\text{e}c^2\,\frac{\nu_{\rm syn}}{\nu_{\rm L}}\, n_\text{e}\left(\gamma_\text{e} = \sqrt{\frac{\nu_{\rm syn}}{\nu_{\rm L}}}\right)\,V\,,
\end{equation}
where $V$ denotes the volume of the emission site.
According to Eq.~(\ref{eq:syn_emis}) the corresponding synchrotron luminosity can be approximated by 
\begin{equation}
    \nu_{\rm syn}L_{\nu_{\rm syn}} \simeq \frac{1}{2}\,\left(\frac{\nu_{\rm syn}}{\nu_{\rm L}}\right)^{1/2}\,P_{\rm syn}\left(\gamma_{\rm e}=\sqrt{\frac{\nu_{\rm syn}}{\nu_{\rm L}}}\right)\,n_{\rm e}\left(\gamma_{\rm e}=\sqrt{\frac{\nu_{\rm syn}}{\nu_{\rm L}}}\right)\,\,V\,,
\end{equation}
so that the total electron energy dependent on the differential synchrotron luminosity at $\nu_{\rm syn}$ yields
\begin{equation}\label{Ee2}
    E_\text{e} \simeq c_{\rm syn}\,B^{-3/2}\,\nu_{\rm syn}L_{\nu_{\rm syn}}\,\quad\text{with}\quad c_{\rm syn}=\frac{12\pi m_\text{e}c}{\sigma_\text{T}}\left(\frac{e}{2\pi m_\text{e}c\nu_{\rm syn}}\right)^{1/2}\,.
\end{equation}
With a magnetic field energy of $E_{\rm B}=V\,U_{\rm B}$, the total energy within the emission site is given by
\begin{equation}
E_\text{tot} = E_\text{e} + E_\text{p} + E_\text{B} = (1+k)c_{\rm syn}\,B^{-3/2}\,\nu_{\rm syn}L_{\nu_{\rm syn}} + \frac{B^2\,r_\text{k}^3}{6}\,,
\end{equation}
where the heavy particle energy $E_{\rm p}$ is $k$ times the electron energy of those electrons that predominantly contribute to $L_{\nu_{\rm syn}}$. Its value depends on the acceleration mechanism of relativistic particles, but in general it is expected that $k\gg 1$. Adopting that the total energy is at its minimum, i.e. $\partial E_{\rm tot}/\partial B=0$, we obtain a magnetic field strength of
\begin{equation}
    B_\text{eq} = (4.5)^{2/7}(1+k)^{2/7}c_{\rm syn}^{2/7}r_\text{k}^{-6/7}\left(\nu_{\rm syn}L_{\nu_{\rm syn}}\right)^{2/7}
\label{eq:eqMag}
\end{equation}
for which the magnetic field energy is about equal to the total particle energy.

\subsection{Hadronic model}\label{hadr}

In the case of an hadronic origin of the observed $\gamma$-ray luminosity, there is no robust low-energy counterpart, such as in the leptonic scenario, that can be used to constrain the energy distribution of the CR protons. 
Moreover, $\gamma$-rays can in principle be produced by photomeson or hadronic pion production, i.e. by inelastic interactions with some target gas and target photon field, respectively. 

In case of photomeson production, the corresponding photon emissivity is given by \citep{Dermer+2009_book}
%But since the typical density of the ambient gas at distances $\gtrsim 10\,\text{pc}$ is small compared to the photon targets, the most promising $\gamma$-ray production process is photomeson production. Its photon emissivity is given by \citep{Dermer+2009_book}
\begin{equation}
   \tilde{\epsilon}_{\pi\gamma}\left(E_{\gamma}\right) = \frac{c\,\zeta_\gamma\, \sigma_{\pi\gamma}^{s,m}}{64\pi^2\chi_\gamma m_\text{p}c^2}\,\frac{n_\text{p}\left(\bar{\gamma}_\text{p}\right)}{\bar{\gamma}_\text{p}^2}\int_{\frac{\varepsilon_{l}^{\prime}}{2\bar{\gamma}_\text{p}}}^{\infty} \diff \varepsilon\frac{n_\text{ph}(\varepsilon)}{\varepsilon^2}f\left(\bar{\gamma}_\text{p},\varepsilon\right)\,,
\end{equation}
where $n_\text{ph}(\varepsilon)$ denotes the differential photon number density at a dimensionless photon energy $\epsilon=h\nu/m_{\rm e}c^2$.
Moreover, the function $f(\bar{\gamma}_\text{p},\varepsilon) \equiv \left\{\big[\min\left(2\bar{\gamma}_\text{p} \varepsilon, \varepsilon_u^{\prime}\right)\big]^2-\varepsilon_{l}^{\prime 2}\right\}\,$ accounts for single pion production, where $\sigma^s_{\pi\gamma}$ = 340 $\mu$b, ${\varepsilon}_l^{\prime}$ = 390, ${\varepsilon}_u^{\prime}$ = 980, as well as multi pion production, where $\sigma^m_{\pi\gamma}$ = 120 $\mu$b, ${\varepsilon}_l^{\prime}$ = 980, ${\varepsilon}_u^{\prime} \longrightarrow \infty$. Further, $\zeta_\gamma^s=1$ and $\zeta_\gamma^m=2$, respectively, denote the multiplicity in case of single and multi pion production respectively, and $\chi_\gamma^{m,s}=0.1$ is the mean fractional energy of the produced $\gamma$-ray compared to the incident primary proton. In addition, $n_\text{p}(\bar{\gamma}_\text{p})$ is the differential intensity of CR protons with a Lorentz factor $\bar{\gamma}_\text{p} \equiv (E_{\gamma})/(\chi_\gamma m_\text{p}c^2)$.

\begin{figure}[ht]
\centering
    \includegraphics[width=0.48\textwidth]{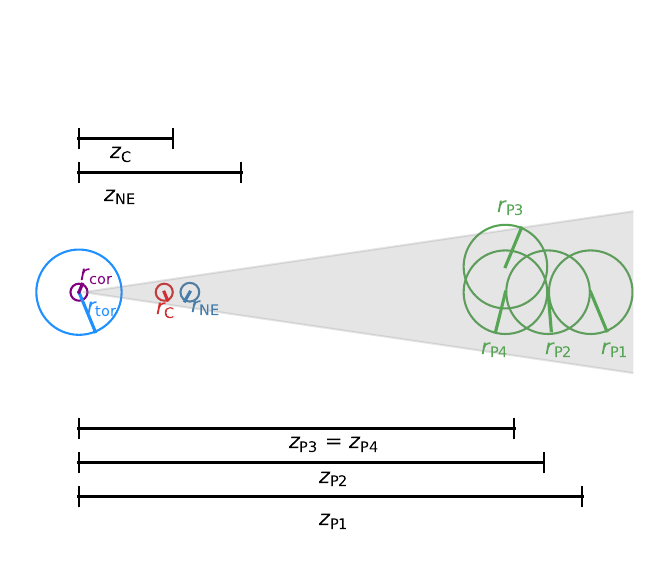}
        \caption{Sketch of the environment, with the distances not in scale ($r_\text{cor} \sim 10^{-4}$ $\rm{pc}$ and $r_\text{tor} = 3.6$ $\rm{pc}$). The radii $r$ of the knots and respective distances $z$ from the core are summarised in Table \ref{em_sites}. }
        \label{backfields1}
\end{figure}

\begin{figure}[ht]
\centering
    \includegraphics[width=0.48\textwidth]{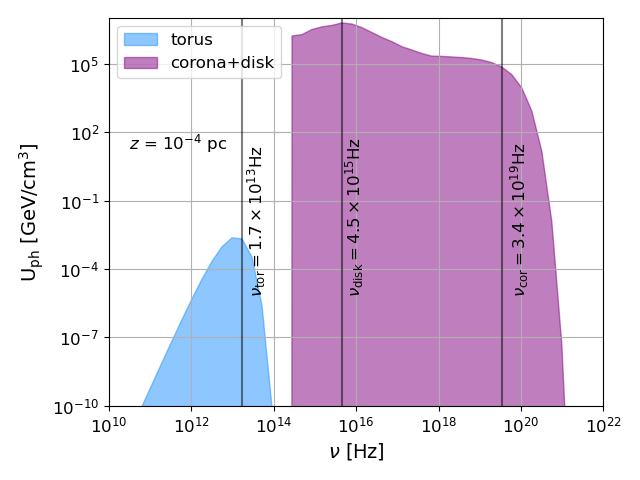}
        \caption{Spectral distribution of the energy density within the torus and corona+disk environment, computed at a distance of $10^{-4}$ pc. The torus field has its peak at a characteristic frequency of $\nu_\text{tor} = 1.7\times 10^{13} \text{Hz}$, the disk at a frequency of $\nu_\text{disk} = 4.5\times 10^{15} \text{Hz}$ and the corona at a frequency of $\nu_\text{cor} = 3.4\times 10^{19} \text{Hz}$.}
        \label{backfields1_}
\end{figure}

In case of hadronic pion production, the corresponding photon emissivity is given by \citep{Schlickeiser2002_book}
\begin{equation}\begin{split}
   \tilde{\epsilon}_{\rm pp}\left(E_{\gamma}\right) &= \frac{2c\,n_{\rm gas}}{3\pi m_\pi c^2}\,\int_{E_{\gamma}+ \frac{m_\pi^2 c^4}{4E_{\gamma}}}^{m_\pi c^2\gamma_{\rm p,max}^{3/4}} \diff E_\pi \, \zeta\sigma_{\rm pp}\left(\left(\frac{E_\pi}{m_\pi c^2}\right)^{4/3}\right)\\
   &\quad \times \left(\frac{E_\pi}{m_\pi c^2}\right)^{1/3}\,n_{\rm p}\left(\left(\frac{E_\pi}{m_\pi c^2}\right)^{4/3}\right)\,\left[E_\pi^2-m_\pi^2c^4 \right]^{-1/2},
\end{split}\end{equation}

\begin{figure}[!ht]
         \includegraphics[width=0.49\textwidth]{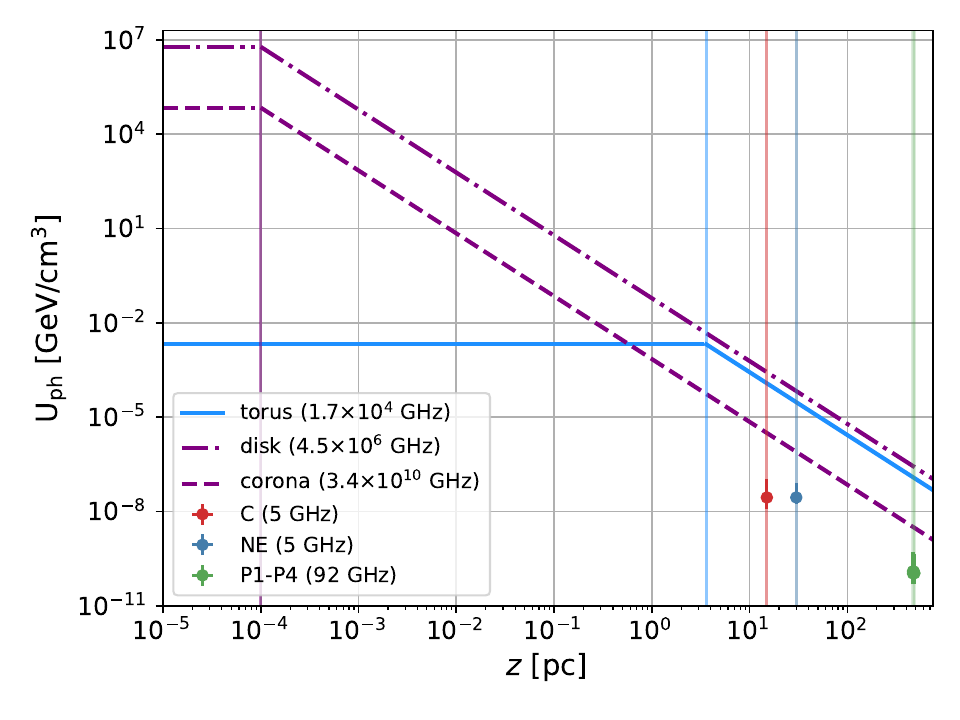}
        \caption{The differential photon energy densities at their characteristic frequencies. The data points for knot C and NE refer to the internal background photon energy densities according to \cite{Gallimore+2004} (at $5\,\text{GHz}$) and the ones for P1-P4 refer to the data from \cite{Michiyama_2022} (at $92\,\text{GHz}$). For the conversion of the radio data of the six knots knot radii of $r_\text{k,1}=0.2\pm 0.1\,\text{pc}$, $r_\text{k,2}=0.3\pm 0.15\,\text{pc}$ and $r_\text{k,3}=3.5\pm 1.75\,\text{pc}$, respectively, are supposed.}
        \label{backfields2}
\end{figure}

\begin{figure}[!ht]
\includegraphics[width=0.49\textwidth]{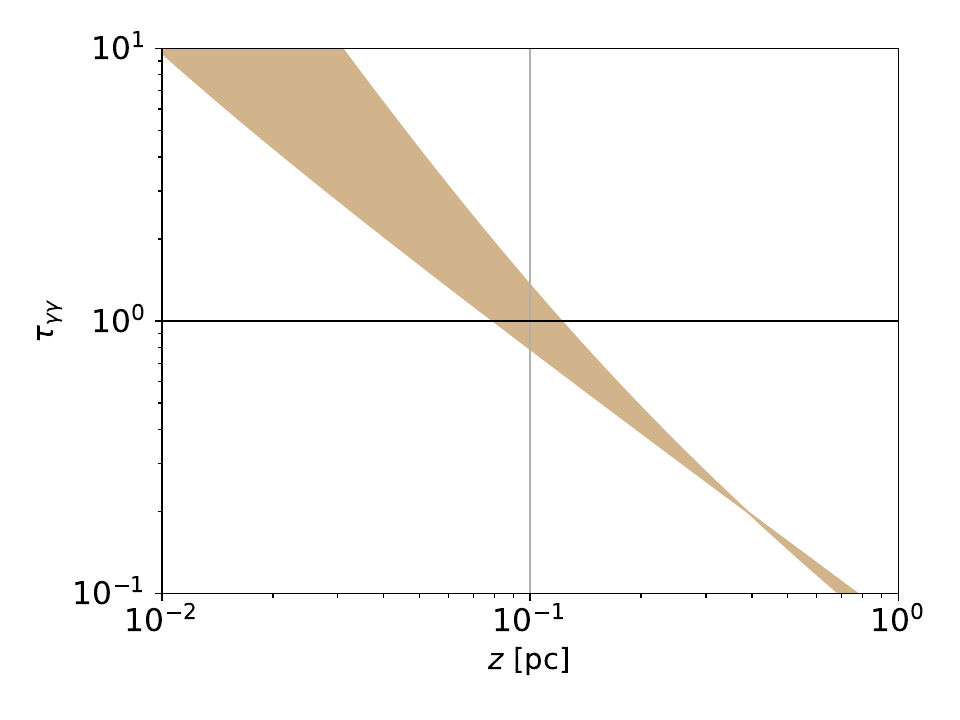}
\caption{Optical thickness $\tau_{\gamma\gamma}$ at 17 GeV dependent on the distance $z$ of the knot with respect to the central engine. Here, the absorption coefficient $\alpha_{\gamma\gamma}$ is calculated according to \cite{Dermer+2009_book} and green shaded range corresponds to the uncertainty of the knot radius evolution (see e.g.\ \cite{Zacharias_2022} for more details).}
\label{backfields2_}
\end{figure}

where $n_{\rm gas}$ denotes the constant target gas density. Moreover, the product of the total cross section $\sigma_{\rm pp}$ and the multiplicity $\zeta$ can be approximated at $\gamma_{\rm p}\geq 1.75$ by $\zeta\sigma_{\rm pp}(\gamma_{\rm p})=8.12\times 10^{-27}\,(\gamma_{\rm p}-1)^{0.53}\,\text{cm}^2$. Note, that at $\gamma_{\rm p}< 1.75$ the steep decline of the inclusive cross section $\zeta\sigma_{\rm pp}$ yields a strong suppression of the emissivity at sub-GeV energies.

Using a differential proton number density $n_\text{p}(\gamma_\text{p}) = n_0\left(\gamma_\text{p}/\gamma_{0,\text{p}}\right)^{-q_\text{p}}$ for $\gamma_{\text{p},\text{min}}<\gamma_\text{p}<\gamma_{\text{p},\text{max}}$, the associated total relativistic proton luminosity, $L_\text{p}\equiv 4\pi r_{\rm k}^2v_{\rm b}\,m_{\rm p}c^2\,\int d\gamma_\text{p}\,\gamma_\text{p}\,n_\text{p}(\gamma_\text{p})$, cannot exceed the available jet power $P_\text{jet}$, i.e. $L_\text{p}=f_{\rm jet}\,P_\text{jet}$ with $f_{\rm jet}<1$.

In this work, a homogeneous spherically symmetric emission site with a radius $r_\text{k}$ is adopted, so that the $\gamma$-ray luminosity at a certain energy $E_\gamma$ yields
\begin{equation}\label{Lgamma_pgamma}
    \nu_{\pi\gamma} L_{\nu_{\pi\gamma}} = A_{\pi\gamma}\, E_\gamma^2\, \frac{c}{v_{\rm b}}\,r_{\rm k}\,f_{\rm jet}P_\text{jet}\, \bar{\gamma}_\text{p}^{-q_\text{p}-2}\,\int_{\frac{\varepsilon_{l}^{\prime}}{2\bar{\gamma}_\text{p}}}^{\infty} {  \diff\varepsilon \frac{n_\text{ph}(\varepsilon)}{\varepsilon^2}f(\bar{\gamma}_\text{p},\varepsilon)}
  %  A_{\pi\gamma}\,\frac{f_{\rm jet}\,P_\text{jet}\,\,(2-q_\text{p})\,\bar{\gamma}_\text{p}^{-q_\text{p}-2}}{\gamma_{\text{p},\text{max}}^{2-q_\text{p}}-\gamma_{\text{p},\text{min}}^{2-q_\text{p}}}\,\int_{\frac{\varepsilon_{l}^{\prime}}{2\bar{\gamma}_\text{p}}}^{\infty} {  \diff\varepsilon \frac{n_\text{ph}(\varepsilon)}{\varepsilon^2}f(\bar{\gamma}_\text{p},\varepsilon)}
\end{equation}
and 
\begin{equation}\label{Lgamma_pp}
\begin{split}
    \nu_{\rm pp} L_{\nu_{\rm pp}} &= A_{\rm pp}\, E_\gamma^2\,\frac{c}{v_{\rm b}} \,\left( \frac{n_{\rm gas}\,r_{\rm k}}{\text{cm}^{-2}} \right)\,f_{\rm jet}P_{\rm jet}\,\int_{E_\gamma+ \frac{m_\pi^2 c^4}{4E_\gamma}}^{m_\pi c^2\gamma_{\rm p,max}^{3/4}} \diff E_\pi \\
    &\quad \times \left( \frac{E_\pi}{m_\pi c^2}\right)^{-\frac{4q_{\rm p}-1}{3}}\,\left[ \left( \frac{E_\pi}{m_\pi c^2}\right)^{\frac{4}{3}}-1 \right]\,\left[E_\pi^2-m_\pi^2c^4 \right]^{-\frac{1}{2}}\,,
    \end{split}
\end{equation}
respectively, with
\begin{equation}
\begin{split}
A_{\pi\gamma}&=\frac{\zeta_\gamma\, \sigma_{\pi\gamma}^{s,m}\,(2-q_\text{p})}{48\pi \,m_{\rm p}^2c^4\,\chi_\gamma\,(\gamma_{\text{p},\text{max}}^{2-q_\text{p}}-\gamma_{\text{p},\text{min}}^{2-q_\text{p}})}\quad\text{and}\\
A_{\rm pp}&=\frac{2.89\times10^{-26}\,(2-q_{\rm p})}{m_\pi m_p c^4 \,(\gamma_{\text{p},\text{max}}^{2-q_\text{p}}-\gamma_{\text{p},\text{min}}^{2-q_\text{p}})}\,.
\end{split}
\end{equation}
Dependent on the present target densities $n_{\rm gas}$ and $n_{\rm ph}$, respectively, the hadronic $\gamma$-ray production is typically dominated by one of these processes.
%
%$A_{\pi\gamma} \equiv \zeta_\gamma\, c\, \sigma_{\pi\gamma}^{s,m}/\left(16\pi\chi_\gamma m_\text{p}c^2\right)$
%
%
%
%
%
%
\subsection{Photon target}

As shown in Fig.~\ref{backfields1_} the most energetic target photon densities result from the torus, which is located at a distance of about $(1-10)\,\text{pc}$ from the BH, and the accretion disk (in the UV band) and the surrounding corona (where these photons are comptonised into the X-ray band), whereof the latter expands up to distances of about 100 Schwarzschild radii \citep{Murase2022}, which is about $10^{-4}\,\text{pc}$ in the case of NGC\,1068. Here, the torus with an average temperature of $T_\text{tor} \sim 150$ K \citep{lopez2018emission} is modelled by a grey body spectrum with
\begin{equation}
    \nu L_\nu^{\text{tor}} = \frac{L_\text{bol}}{6}\left(\frac{h\nu(1+z)}{k_\text{B}T_\text{tor}}\right)^4\,\exp\left(-\frac{h\nu(1+z)}{k_\text{B}T_\text{tor}}\right)
\end{equation}
where the prefactor of $1/6$ ensures that the integral over $L_\nu^{\text{tor}}$ yields the bolometric luminosity of
$L_{\rm bol}=2.46\times 10^{44}\,\text{erg/s}$, as provided by the so-called CLUMPY torus model by \cite{lopez2018emission}. Here, an outer radius of $r_{\rm tor}=3.6\,\text{pc}$ has been adopted\footnote{Under consideration of the different assumptions on the distance of NGC\,1068.}. Note that the alternative scenario of a smooth torus yields a photon density that is about a factor of 0.5 times smaller.
The term $T_\text{tor}$ denotes the temperature of the dusty torus, and we consider it to be $T_\text{tor} = 150 \text{K}$.

\begin{comment}
\begin{figure}
         \includegraphics[width=0.49\textwidth]{density_dist.pdf}
          \includegraphics[width=0.49\textwidth]{opt_thick.pdf}
        \caption{\textit{Left}: The differential photon energy densities at their characteristic frequencies. The data points for knot C and NE refer to the internal background photon energy densities according to \cite{Gallimore+2004} (at $5\,\text{GHz}$) and the ones for P1-P4 refer to the data from \cite{Michiyama_2022} (at $92\,\text{GHz}$). For the conversion of the radio data of the six knots a knot radius of $r_\text{b,1}=2.5\pm 1\,\text{pc}$, $r_\text{b,2}=4.6\pm 1.8\,\text{pc}$ and $r_\text{b,3}=3.5\pm 1.4\,\text{pc}$, respectively, is supposed. \textit{Right}: Optical thickness $\tau_{\gamma\gamma}$ at 17 GeV dependent on the distance $d$ of the knot with respect to the central engine. Here, the absorption coefficient $\alpha_{\gamma\gamma}$ is calculated according to \cite{Dermer+2009_book} and green shaded range corresponds to the uncertainty of the knot radius evolution (see e.g. \cite{Zacharias_2022} for more details).}
        \label{backfields2}
\end{figure}
\end{comment}

The photon field of the accretion disk is modelled by an empirical model by \cite{Ho2008}, where the UV spectrum is determined from the given Eddington ratio $\lambda_\text{Edd} \equiv L_\text{bol}/L_\text{Edd}\sim 1.4$ of NGC\,1068. The subsequent coronal X-ray spectrum is modelled by a power-law with an exponential cutoff. Here, the photon index $\Gamma_X$ is correlated to $\lambda_\text{Edd}$ through $\Gamma_X \approx 0.167 \times \log(L_\text{Edd})$ + 2 and the cutoff energy is given by $E_{\text{X,cut}} \sim [-74\log(\lambda_\text{Edd}) + 1.5 \times 10^2]$ keV \citep{a5:Murase+2020}. At low energies, we cut off the spectrum for energies lower than 1 eV, since the infrared photons at these energies originate predominantly from the previously described torus at significantly larger distances from the BH. 
%Note that this emission from the nuclear region of NGC\,1068 gets reprocessed by the broad line region yielding a slightly attenuated photon density \textbf{(@Xavier/Ralf-Jürgen: Can this be quantified somehow better?!)} at distances of the considered emission sites. However, at most frequencies this attenuation is rather small \textbf{(@Xavier/Ralf-Jürgen: Can this be quantified somehow better?!)}, so that we will neglect the impact of the broad line region in the following.
Note that this emission from the nuclear region of NGC\,1068 gets reprocessed/ scattered by the clouds in the broad line region, so that a fraction $f_{\rm diff}\ll 1$ of the central luminosity $\nu_0 L_{\nu_0}$ forms a diffuse radiation field at a distance from the BH of $z\sim 0.1\,\text{pc}$. However, at the considered emission sites, where $z\gg 1\,\text{pc}$, this diffuse photon field appears again as a point source yielding a photon energy density in the reference frame of the knot of \citep{Dermer+2009_book} 
\begin{equation}
    U_{\rm trg} = \frac{\nu_0 L_{\nu_0}}{4\pi z^2c}\,\left[ \Gamma_{\rm b}^2(1+\beta_{\rm b})^2 
 \right]^{-1}\,,
 \label{eq:exPhotTarget}
\end{equation}
where $\beta_{\rm b}c$ denotes the velocity of the knot and $\Gamma_{\rm b}$ its corresponding Lorentz factor. Note that at small distances from the isotropic photon target, the previous point source assumption is no longer valid and details of the geometry of the isotropic radiation field have to be taken into account to obtain the photon energy density in the reference frame of the knot, see e.g.~\cite{Sikora+1994} for more details. But since all of the considered emission sites show non-relativistic velocities 
with $\beta_{\rm b} \equiv v_{\rm b}/c \lesssim 0.05$ \citep{roy2000slow}, we can neglect these relativistic beaming effects in the following. 
%a slightly attenuated photon density \textbf{(@Xavier/Ralf-Jürgen: Can this be quantified somehow better?!)} at distances of the considered emission sites. However, at most frequencies this attenuation is rather small \textbf{(@Xavier/Ralf-Jürgen: Can this be quantified somehow better?!)}, so that we will neglect the impact of the broad line region in the following. 
%Moreover, all of the considered emission sites show non-relativistic velocities with $\beta_{\rm b} \equiv v_{\rm b}/c \lesssim 0.05$ \citep{roy2000slow}. Therefore, the Doppler factor $\delta$ is close to unity, and the impact of Doppler boosting of the target photon fields into the reference frame of the emission site is negligible. Hence, the photon density $U_{\rm trg}$ at a distance $(d-r_{\rm ch})$ from the considered emission region (i.e. either the torus or the corona) is independent of its adopted characteristic size $r_{\rm ch}\ll d$ yielding \textbf{(@Xavier: is the doppler factor as introduced here correct?)}
%\begin{equation}
%    U_{\rm trg} = \frac{\nu_0 L_{\nu_0}}{4\pi r_\text{ch}^2c} \left(\frac{r_\text{ch}}{d}\right)^{2}\delta^4 = \frac{\nu_0 L_{\nu_0}}{4\pi d^2c}\delta^4\,.
%\end{equation}
%Hereby, 
Note that Eq.~(\ref{eq:exPhotTarget}) can also be used for the isotropic photon field from the torus, with a luminosity $\nu_0 L_{\nu_0}$, if we adopt that $z\gg r_{\rm tor}$ and the spatial extension of the torus is not orders of magnitudes smaller than $r_{\rm tor}$. As shown in Fig.~\ref{backfields2} the photon energy density from the corona+disk environment is at all distances larger than the differential photon density of the torus at its peak frequency of $\nu_\text{tor} = 1.7\times 10^{13}$ Hz. Moreover, it is shown in Fig.~\ref{backfields2} for the considered emission sites with a knot radius $r_{\rm k}$ as well as an observed luminosity $\nu_{\rm obs}L_{\rm obs}$ in the radio band, as given in Table \ref{em_sites}, that the internal photon field\footnote{With internal we are referring to the photon density that is produced within the considered knot structures, in contrast to the previously discussed external fields.} energy density at $\nu_{\rm obs}$ yields
\begin{equation}\label{usyn}
    U_\text{trg}^{\rm (int)} = \frac{\nu_{\rm obs}L_{\rm obs}}{4\pi r_{\rm k}^2 c} \ll U_{\rm trg}\,.
\end{equation}
This also holds for the considered emission sites if we account for the observed spectral behaviour $\alpha$ and compare the energy densities at the characteristic frequencies of the torus or the disk.  
Hence, in the following only the external photon fields are taken into account, as the dominant photon target to generate $\gamma$-rays up to some tens of GeV. 

In general, the $\gamma$-ray luminosity, according to Eq.~(\ref{eq:LumRatio0}), (\ref{Lgamma_pgamma}) and (\ref{Lgamma_pp}), scales linearly with the present target photon density, if absorption via $\gamma\gamma$ pair production (with an absorption coefficient $\alpha_{\gamma\gamma}$) can be neglected. Fig.~\ref{backfields2_} shows that this is actually a proper presumption as even at the highest considered $\gamma$-ray energies the corresponding optical thickness $\tau_{\gamma\gamma}\equiv  \alpha_{\gamma\gamma}\,r_{\rm k}\ll 1$ for the considered radio knots. Moreover, it illustrates that dependent on the supposed knot radius evolution the observed $\gamma$-rays at a few $\times 10\,\text{GeV}$ cannot be explained by alternative emission sites that are closer than about a few $\times 0.1\,\text{pc}$ to the central nucleus.

\section{Results}\label{sec:results}
\subsection{Leptonic Scenario}\label{sec:leptonic_scenario}

\begin{figure}[htb]
\centering
         \includegraphics[width=0.5\textwidth]{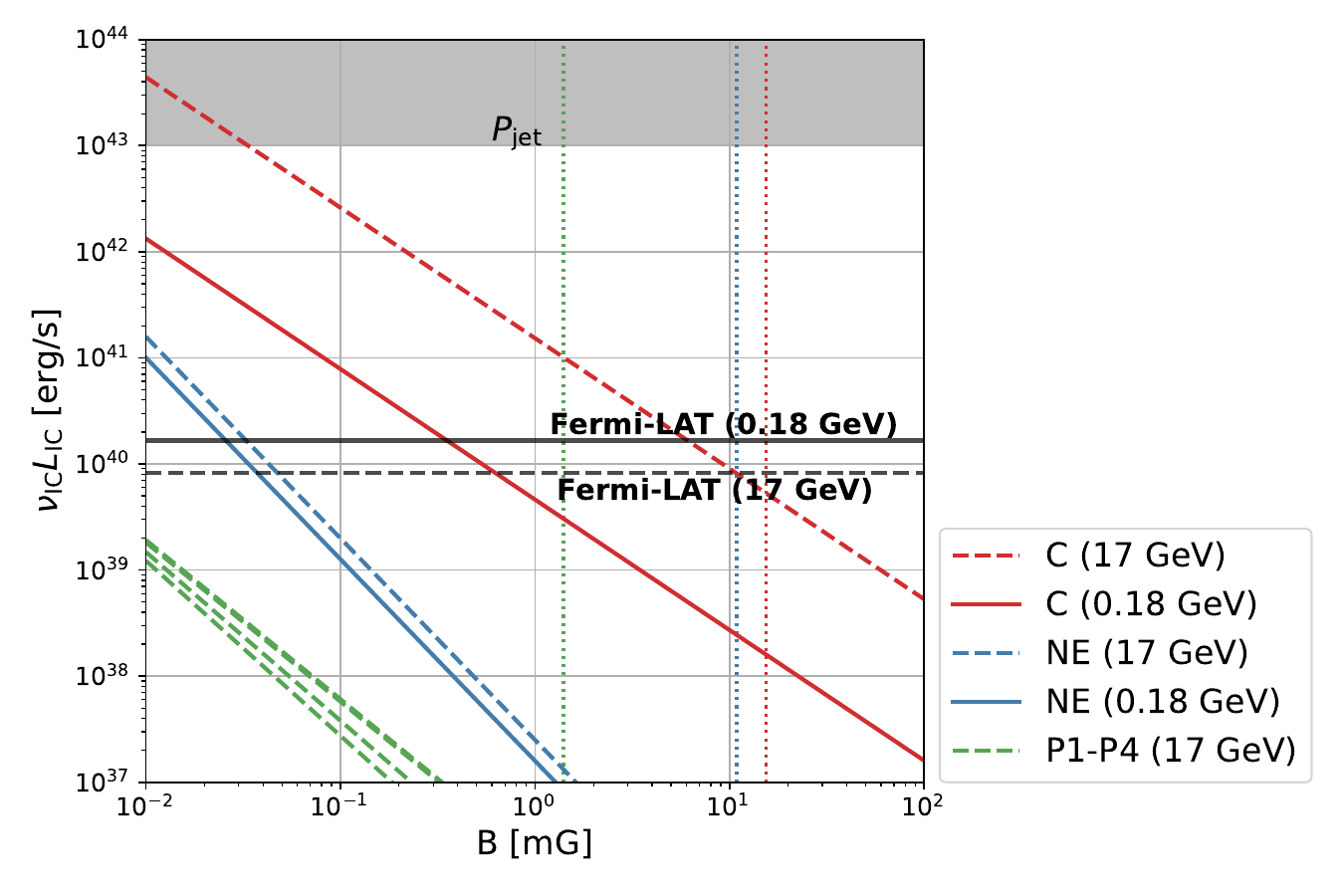}
        \caption{Produced luminosity via IC process, according to Eq.~(\ref{eq:LumRatio0}). The dotted lines indicate the values of the magnetic field according to Eq.~(\ref{eq:eqMag}). The grey shaded band excludes all values higher than $P_{\rm jet}$.}       
        \label{leptonic_results}
\end{figure}

In the following, we determine the $\gamma$-ray luminosity at the low and high-energy end of the recently updated Fermi-LAT flux \citep{Abdollahi_2022}, i.e. at $\nu_{\rm IC,1}=0.18\,\text{GeV}/h$ (luminosity of $1.7\times 10^{40}$ erg/s) and $\nu_{\rm IC,2}=17\,\text{GeV}/h$ (luminosity of $8.2\times 10^{39}$ erg/s). Hereby, we use Eq.~(\ref{eq:LumRatio0}) to determine the $\gamma$-ray luminosity  from IC scattering in the Thomson limit, so that%\textbf{(@Xavier: is the doppler factor as introduced here correct?)}
\begin{equation}
     \nu_{\rm IC}L_{ \nu_{\rm IC}} = 2 \left[ \frac{3}{8\pi}\,\frac{\nu_{\rm IC}/\nu_{0}}{\nu_{\rm syn}}\,\frac{e}{m_{\rm e} c} \right]^{\frac{3-q_{\rm e}}{2}}\,\frac{\nu_0 L_{\nu_0}}{z^2c}\,B^{-\frac{1+q_{\rm e}}{2}}\,\nu_{\rm syn}L_{\nu_{\rm syn}},
     \label{eq:LumRatio0_2}
\end{equation}
where $\nu_{\rm syn}L_{\nu_{\rm syn}}=\nu_{\rm obs}L_{\nu_{\rm obs}}$ at $\nu_{\rm syn}=\nu_{\rm obs}$ as given in Table \ref{em_sites}. A maximal $\gamma$-ray luminosity is obtained for a dense target energy distribution at $\nu_0$ as well as a low value of $\nu_0$ in case of $q_{\rm e}<3$. In case of optically thin synchrotron emission at $\nu_{\rm syn}$ according to $L_{\nu_{\rm syn}}\propto \nu_{\rm syn}^{-\alpha}$ the spectral behavior of the relativistic electron distribution can be determined by $q_\text{e} = 2\alpha + 1$. Note that this spectral behavior can in principle become softer towards higher energies due to the impact of synchrotron or IC losses. But without these softening effects the spectral behavior as given in Table \ref{em_sites} indicates that the torus with a spectral luminosity of $\nu_0 L_{\nu_0}=1.72\times 10^{44}\rm {erg/s}$ at $\nu_0=\nu_\text{tor} = 1.7\times 10^{13}\,\text{Hz}$ is the dominant photon target. 
As shown in Fig.~\ref{leptonic_results} the emission sites C is at $\nu_{\rm IC,2}=17\,\text{GeV}/h$ able to exceed the observed $\gamma$-ray luminosity for the whole range of reasonable magnetic field strength values, such as its $B_{\rm eq}$ value (for a proton-to-electron energy ratio of $k=100$) that is indicated by the red dotted line. Note that for all emission sites the values of $B_{\rm eq}$ are rather large. 
Comparing the results at $0.18$ and $17\,\text{GeV}$, it is clear that the adopted spectral behavior of the relativistic electrons, that has been derived from the radio emission, is too hard with respect to the observed spectral behavior in the $\gamma$-ray band. Thus, this emission site has in principle enough total energy to produce the observed $\gamma$-rays above about a few GeV, if electrons can be accelerated up to $\gamma_{\rm e}\simeq \sqrt{3\nu_{\gamma}/4\nu_{\rm tor}}\simeq 4\times 10^5$. But to explain also the sub-GeV $\gamma$-ray emission, a magnetic field strength of about $0.5\,\text{mG}$ is needed, which is about an order of magnitude smaller than what is expected from the minimal total energy estimate. Moreover, a significant softening of the spectral behavior of the CR electrons to an index of $q_{\rm e}\sim 3.3$ at $\gamma_{\rm e}\simeq \sqrt{3\nu_{\rm IC,1}/4\nu_{\rm tor}}= 4\times 10^4$ is needed
%(as further discussed in Sect.~\ref{sec:disc&conc}) 
to agree with the spectral behavior of the observed $\gamma$-ray data. 
\begin{figure}[htb]
\centering
\includegraphics[height=6cm]{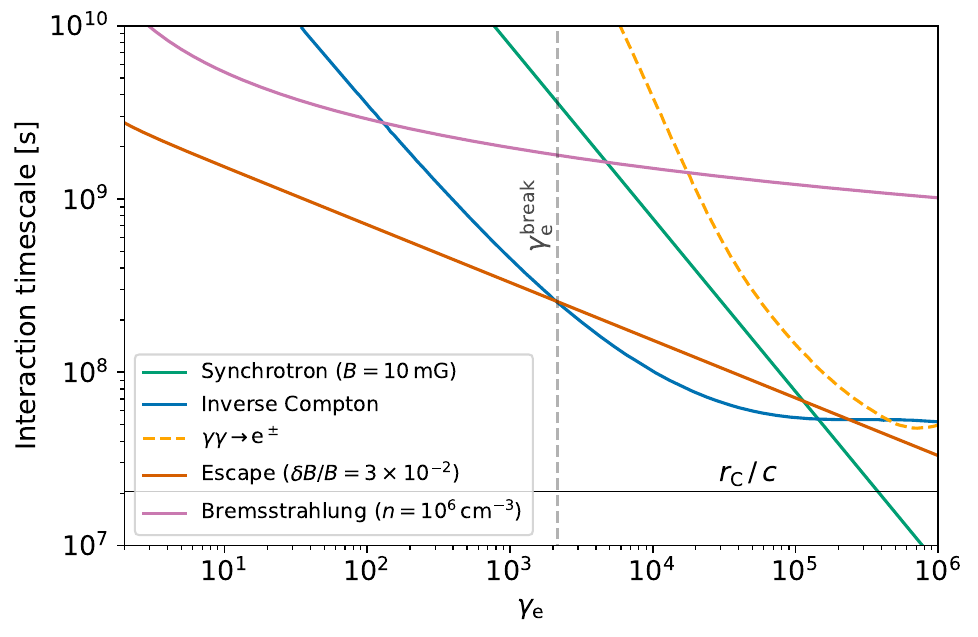}
    \caption{Interaction timescales for the relevant radiative processes in C knot. The horizontal black line represents the size of the region based on radio observations. We can see that inverse Compton cooling (blue curve) becomes efficient for electron energies  $E_\mathrm{e}>1.1~\mathrm{GeV}$, as denoted by the vertical line, leading to a cooling break in the electron spectrum.}
    \label{timescales_component_c}
\end{figure}
%
%%%NEW:
Hence, in addition to the low magnetic field strength, the data also shows the need for a strong cooling break in the electron spectrum at about 1\,\text{GeV}. But as shown in Fig.~\ref{timescales_component_c} a break at about these energies can only occur if the escape of CR electrons is delayed by at least an order of magnitude with respect to the free streaming condition which is determined by the light-crossing time of the knot, suggested to be about 0.2~pc by VLBA observations (cf. Tab.~\ref{em_sites}). This can in principle be realized by diffusive escape from the knot, which however, already softens the initial spectral behavior so that only a minor softening by $\Delta q_{\rm e}\simeq 0.6$ emerges at the transition from diffusive escape to IC losses. 
Moreover, we can see that synchrotron emission (green line) is comparatively inefficient, since the energy density of the magnetic field is much lower than that of the external thermal fields. That also applies to relativistic bremsstrahlung losses (purple line) even for the case of a dense target plasma. For completeness, we also show the optical thickness to photon-photon pair production (dashed orange curve), which is inefficient up to photon energies of about 30~GeV, as shown previously in Fig.~\ref{backfields2_}.
Thus, all potential interaction processes in knot C are not able to explain the necessary softening of the spectrum.
%As shown in Fig.~\ref{timescales_component_c} this can in principle be realized by diffusive escape from the knot, if the characteristics of the turbulent magnetic fields are appropriately. In any case, it is clear that the escape of CR electrons needs to be delayed by at least an order of magnitude with respect to the free streaming condition which is determined by the light-crossing time of the knot, suggested to be about 0.2~pc by VLBA observations (cf. Tab.~\ref{em_sites}) 
%Moreover, we can see that synchrotron emission (green line) is comparatively inefficient, since the energy density of the magnetic field is much lower than that of the external thermal fields. For completeness, we also show the optical thickness to photon-photon pair production (dashed orange curve), which is inefficient up to photon energies of about 30~GeV, as shown previously in Fig.~\ref{backfields2_}.
%For a sufficiently dense target plasma within the knot C it is also possible that low energy electrons are cooled dominantly by relativistic bremsstrahlung losses. In that case a plasma density of about \textbf{???} is needed to obtain the spectral break at the right energy. 
%%%%%
A significant $\gamma$-ray contribution from the NE knot is only obtained for a magnetic field strength $\ll 0.1\,\text{mG}$. Further, it can be excluded that the observed $\gamma$-rays originate from the head of the jet (i.e. P1-4).

\begin{table}[ht]
  \caption{Properties for the different emission sites that have been observed in the radio band by \cite{Gallimore+2004} (for C and NE) and \cite{Michiyama_2022} (for P1-4) as well as the expected magnetic field strength according to Eq.~(\ref{eq:eqMag}).}
  \centering 
  \begin{threeparttable}
  \resizebox{8.8cm}{!}{
    \begin{tabular}{lccccccc} 
          & $z$ & $r_\text{k}$  & $\nu_\text{obs}$  & $\nu_\text{obs}L_{\nu_\text{obs}}$  & $\alpha$ & $B_{\rm eq}(k=100)$  \\
          & [pc] &  [pc] &  [GHz] &  [$10^{36}$\,\text{erg/s}]  & & [mG] \\
     \midrule    
    C & 15 &   0.2  & 5 & $6.4$ & 0.23 &  15.4 \\
    NE & 30 &   0.3  & 5 & $9.5$ & 0.90 &  10.9 \\
    P1 & 484 &   3.5  & 92 & $7.6$ & 0.50 &  1.40 \\ 
    P2 & 477 &   3.5  & 92 & $8.6$ & 0.59 & 1.40\\ 
    P3 & 468 &   3.5  & 92 & $8.8$ & 0.65 & 1.40 \\ 
    P4 & 468 &   3.5  & 92 & $7.5$ & 0.50 & 1.40\\ 
    \midrule
    \end{tabular}
    }
    \begin{tablenotes}
      \small
      \item \textbf{Notes.} Length scales are determined supposing a source distance of $10.1\,\text{Mpc}$, so that $1^{\prime\prime}=49\,\text{pc}$.
      %We label as C the emission site at 15 pc, NE the one at 30 pc and P1, P2, P3, P4 the 4 knots at 484 pc. The last emission site is indeed composed of 4 knots with similar characteristics. 
      Here, $z$ is the distance of the knot with respect to the central engine,  $r_{\text{k}}$ is the knot radius, $\nu_\text{obs}L_{\nu_\text{obs}}$ the observed radio luminosity at a frequency $\nu_\text{obs}$ and $\alpha$ the spectral index at this frequency.
    \end{tablenotes}
\end{threeparttable}
\label{em_sites}
\end{table}

\subsection{Hadronic Scenario}
Accounting for the external photon targets by the torus, disk and corona with a differential photon number density
\begin{equation}\label{nph}
    n_\text{ph}(\varepsilon) = \frac{1}{4\pi}\,\frac{m_{\rm e}c^2}{(h\nu_0)^2}\,U_{\rm trg}
\end{equation}
at a dimensionless photon energy $\epsilon=h\nu_0/m_{\rm e}c^2$, we determine in the following the resulting $\gamma$-rays according to photomeson production as given by Eq.~(\ref{Lgamma_pgamma}). 
Since the torus only provides photons at energies $\ll 100\,\text{keV}$, these target photons are not energetic enough to provide $\gamma$-rays up to some tens of GeV. However, the X-ray photons from the corona with energies up to few $\times 100\,\text{keV}$ are in principle energetic enough to yield $\gamma$-rays at these energies, but not at sub-GeV energies. Moreover, $\nu_{\pi\gamma} L_{\nu_{\pi\gamma}}$ is---even for the most optimistic scenarios---also at $17\,\text{GeV}$ multiple order of magnitudes below the observations. Hence, we can clearly rule out that the jet produces the observed $\gamma$-rays by a photohadronic scenario. 

\begin{figure}[htb]
\centering
\includegraphics[width=0.48\textwidth]{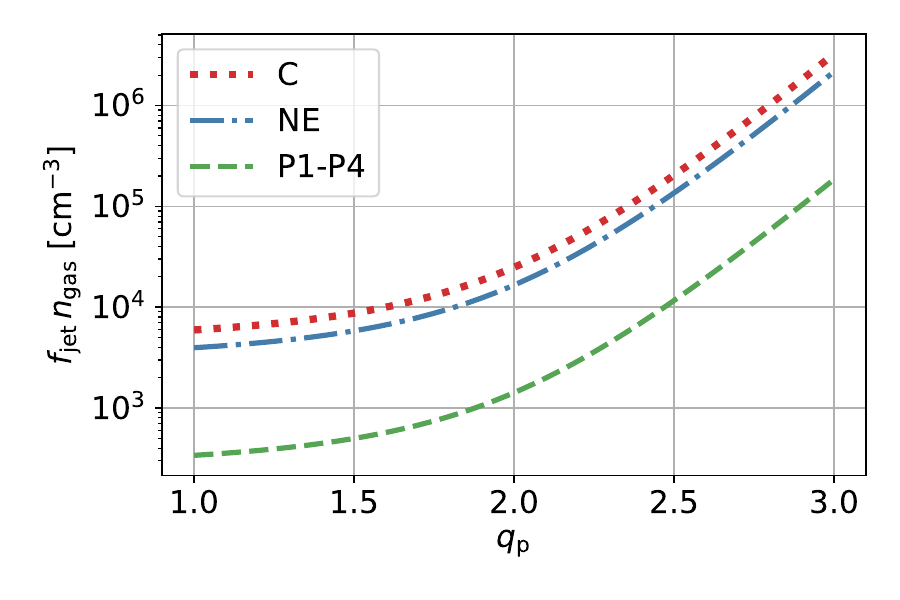}
\caption{The necessary parameter condition for the different knot environments C, NE and P1 to P4 to explain the observed $\gamma$-ray flux of $8.2\times 10^{39}\,\text{erg}\,\text{s}^{-1}$ at $17\,\text{GeV}$ by hadronic pion production. Here a knot velocity of $v_{\rm b}=0.075\,c$ is used as well as $\gamma_{\text{p},\text{min}}=1.$ and $\gamma_{\text{p},\text{max}}=2\times 10^3$ to obtain a cut-off in the resulting $\gamma$-ray flux at about a few $\times10\,\text{GeV}$ as indicated by the observations.}
         %\includegraphics[width=0.6\textwidth]{photomeson_gammaFluxSpec}
        %\caption{$\gamma$-ray flux from photomeson production for $\gamma_{\rm p, max}=200$ and $\gamma_{\rm p, min}=1$ as well as spectral indices in the range $q_{\rm p}\in[1,3]$. Moreover, we used a rather optimistic value of $f_{\rm jet}=0.5$ as well as $P_{\rm jet}=10^{43}\,\text{erg}\,\text{s}^{-1}$.}   
        \label{hadronic_resultPlot}
\end{figure}

In the case of hadronic pion production, the resulting luminosity $\nu_{\rm pp} L_{\rm pp}$ is at a few GeV much closer to the observations than the photohadronic scenario. But due to the fundamental kinetics of the inelastic scattering the observed sub-GeV $\gamma$-rays can still not be explained. In order to match the observed $\gamma$-ray luminosity at $17\,\text{GeV}$, the Fig.~\ref{hadronic_resultPlot} shows the necessary product of target gas density ($n_{\rm gas}$) and the fraction ($f_{\rm jet}$) of the jet power that goes into CR protons dependent on their spectral index ($q_{\rm p}$). Even in the most optimistic scenario of a jet that is completely dominated by the CR protons, i.e. $f_{\rm jet}\simeq 1$, the target gas density needs to be multiple orders of magnitude higher than the average ISM density of about $1\,\text{cm}^{-3}$. But as shown by \cite{garcia2014molecular} dense molecular gas with $n(H_2)\geq 10^{5-6}\,\text{cm}^{-3}$ is present within the circumnuclear disk of NGC\,1068. In particular the knot C seems to impact a giant molecular cloud \citep{Gallimore+2004} not least due to the need of free-free absorption at $1.4\,\text{GHz}$. Thus, if the target gas in the considered knot structures also shows an increased density, the slight excess of the observed $\gamma$-ray flux at about $17\,\text{GeV}$ could actually be a result of the contribution from $\gamma$-rays by hadronic pion production. But in any case there is an additional $\gamma$-ray emitter needed to also explain the observed $\gamma$-rays at sub-GeV energies. 
%\begin{equation}
%    \left( \frac{v_{\rm b}/c}{0.075}\right)^{-1}\,\left( \frac{n_{\rm gas}}{10\,\text{cm}^{-3}}\right)\,\left( \frac{r_{\rm k}}{5\,\text{pc}}\right)\,\left( \frac{f_{\rm jet}P_{\rm jet}}{10^{42}\,\text{erg}\,\text{s}^{-1}}\right)\simeq...
%\end{equation}

%$\gamma$-rays can also not be produced by photomeson production with these coronal X-ray photons. Moreover, the Fig.~\ref{hadronic_resultPlot} shows that also at the highest energies, the resulting flux $\nu_{\pi\gamma} F_{\nu_{\pi\gamma}}=\nu_{\pi\gamma} L_{\nu_{\pi\gamma}}/4\pi D_{\rm L}^2$ is even for the most optimistic scenarios multiple order of magnitudes below the observations. 
%Hence, we can clearly rule, that the jet produces the observed $\gamma$-rays by an hadronic scenario. Moreover, an hadronic origin of the observed $\gamma$-rays at sub-GeV energies can be excluded in general, i.e. independent of the considered emission side, due to the maximal energies of the available target photon fields.

\section{Discussion and conclusions}\label{sec:disc&conc}
In this work, we discuss the possibility that the different radio emission sites, so-called knots, within the jet-like structure of NGC\,1068 can be associated as the origin of the observed $\gamma$-ray flux between about $0.1$ and $100\,\text{GeV}$. 
Hereby, we account for the observed spatial and spectral details of these radio structures and discuss the potential leptonic and hadronic $\gamma$-ray emission scenarios. 

%For relativistic protons within these radio knots, photomeson production with the hard X-ray photons from the inner corona is the most promising scenario. However, due to the limited power of the jet as well as the target photon energy density at these distances, the resulting $\gamma$-ray flux is for all radio knots even for the most optimistic scenarios multiple orders of magnitude smaller than what has been observed.

For relativistic protons within these radio knots, hadronic pion production is the most promising scenario to explain the mild excess of the observed $\gamma$-ray flux at about $17\,\text{GeV}$. For this purpose, it is needed that 
$$\left( \frac{v_{\rm b}/c}{0.075}\right)^{-1}\,\left( \frac{n_{\rm gas}}{1\,\text{cm}^{-3}}\right)\,\left( \frac{r_{\rm k}}{5\,\text{pc}}\right)\,\left( \frac{f_{\rm jet}P_{\rm jet}}{10^{43}\,\text{erg}\,\text{s}^{-1}}\right)\sim 10^{2-3}\,,$$
hence the target gas needs to be significantly denser than in the average ISM. In particular the knot C shows indications of the impact of a dense molecular cloud \citep{Gallimore+2004} that would be sufficient to explain the GeV $\gamma$-rays. 
%Recent observations \citep{garcia2014molecular} indicate such an enhanced molecular gas density within the circumnuclear disk, but to our knowledge this has not been correlated so far with respect to the observed radio knots. 
But in any case, an additional leptonic $\gamma$-ray contribution is needed to account for the observed sub-GeV $\gamma$-rays.

For relativistic electrons, IC scattering with the IR photons from the torus yields the highest flux of $\gamma$-rays based on the expected spectral behavior of the electrons from radio observations. 
Depending on the supposed magnetic field strength within these radio knots we could constrain that the knot C, which
is the closest with respect to the central engine, represents the most likely emission site for the production of the observed $\gamma$-rays.
%%%NEW:
But for this purpose it is necessary that: (i) the magnetic field strength is $\lesssim 1\,\text{mG}$; (ii) the electron spectrum needs to show a break with a strong softening at about an energy of $(1-10)\,\text{GeV}$ to agree with the observed spectral behavior of the $\gamma$-ray flux.
%We show that such a softening results naturally from the dominance of IC losses at energies $\gtrsim 1\,\text{GeV}$. Finally, a best-fit scenario is presented that accounts for the VLBA data of that particular emission site and yields a proper explanation of the observed $\gamma$-rays from the Fermi-LAT. 
%But for this purpose it is necessary that: (i) the magnetic field strength is $\lesssim 1\,\text{mG}$, although significantly higher values are expected from the minimal total energy condition (yielding $B_{\rm eq}\simeq 1\,\text{mG}$); (ii) the electron spectrum needs to show a break with a significant softening at about an energy of $\sim 10\,\text{GeV}$ to agree with the observed spectral behavior of the $\gamma$-ray flux. 
However, we also show that such a strong softening (by $\Delta q_{\rm e}\simeq 1.8$ if at low energies $q_{\rm e}\simeq 1.5$ as indicated from VLBA observations) cannot emerge from the potential interaction processes in knot C. 
%For the characteristic loss timescale $\tau_{\rm loss}\propto \gamma_{\rm e}^r$, it is necessary that $r\sim 0$ becomes $r\lesssim -1$ at about this energy. Hence, there needs to be a transition from e.g.\ non-thermal bremsstrahlung losses towards IC losses, which yields a necessary plasma density on the order of about $10^3\,\text{cm}^{-3}$ \textbf{(still the correct value??)} within knot C. 
%Interestingly this value of the gas density matches by chance the necessary target gas density for the hadronic pion production to produce the mild $\gamma$-ray excess at about $17.2\,\text{GeV}$ as previously discussed.  
In general, these findings are in good agreement with the results from \cite{Lenain+2010}, where however, the previously mentioned conditions (i) and (ii) have been supposed, to explain the $\gamma$-ray data by IC scattering with IR photons from the torus. Nonetheless, these conditions do not hold, because the radius and magnetic field needed to achieve the right $\gamma$-ray luminosity are $\sim$ 0.1 pc and $0.1$ mG. These values are, respectively, an order of magnitude lower and higher than what is observed and expected. Moreover, we do not naturally find a spectral break at an energy of $(1-10)\,\text{GeV}$.
%%%I shift the following sentence into the introduction:
%However, in their work the emission site is considered at a distance from the torus of $65\,\text{pc}$, its radius is set to $r_\text{b} = 7\,\text{pc}$ and a magnetic field of $0.1\,\text{mG}$ is considered. 
Hence, by accounting for the observational constraints by \cite{Gallimore+2004} it is not possible to explain the observed $\gamma$-rays by the radio knot C or any other observed knot structure. 
%The additional contribution from the other radio emission sites is in terms of the leptonic scenario negligible. 
But if one of these sites impacts a dense target gas, they could still provide some hadronic $\gamma$-ray contribution at about $10\, \text{GeV}$.  

One of the key parameters of these results is the strength of the magnetic field: based the minimal total energy condition, as given in Eq.~(\ref{eq:eqMag}), a value of about $(1+k)^{2/7}\,\text{mG}$ is obtained, which however still includes the uncertain proton-to-electron energy ratio $k$. Moreover, this classical estimate can be revised, because it does not account for inhomogeneous magnetic fields, the synchrotron spectrum over the real energy range and the real ratio $k$ of total energies. Considering these factors, it has been shown by \cite{beck2005revised1} that the resulting magnetic field could become slightly smaller by a factor of about $0.86 - 0.99$. Hence, even in the most extreme scenario of a vanishing hadronic particle energy, the field strength $B_{\rm eq}$ that are given in Table \ref{em_sites} would only become smaller by a factor of about $0.2$. 

Further, knot C only manages to produce sufficient $\gamma$-rays due to its non-relativistic bulk motion. In the case of a mildly relativistic motion, i.e. $\Gamma_b\gg 1$, the target photon energy density in the rest frame of the knot would be reduced according to Eq.~(\ref{eq:exPhotTarget}) and the target photon frequency $\nu_0$ would suffer from relativistic Doppler de-boosting which would in total lower the expected IC $\gamma$-ray luminosity by a factor $\Gamma_{\rm b}^{-(1+q_{\rm e})/2}$. 

Apart from the previously discussed spatially resolved radio knots, there could be the chance that the observed $\gamma$-rays are produced much closer to the foot of jet, where radio telescopes cannot resolve individual emission sites. But as shown in Fig.~\ref{backfields2_}, the emission sites already become opaque for $\sim 10\,\text{GeV}$ $\gamma$-rays at distances of about $0.1\,\text{pc}$ due to $\gamma\gamma$ pair production. Therefore, such alternative emission sites could not explain the observed $\gamma$-rays at the highest energies.\newline\newline
\textit{Acknowledgements}. We acknowledge funding from the German Science Foundation DFG, within
the Collaborative Research Center SFB 1491 “Cosmic Interacting Matters - From Source to Signal”.

\bibliography{references}{}

\begin{thebibliography}{}
\expandafter\ifx\csname natexlab\endcsname\relax\def\natexlab#1{#1}\fi
\providecommand{\url}[1]{\href{#1}{#1}}
\providecommand{\dodoi}[1]{doi:~\href{http://doi.org/#1}{\nolinkurl{#1}}}
\providecommand{\doeprint}[1]{\href{http://ascl.net/#1}{\nolinkurl{http://ascl.net/#1}}}
\providecommand{\doarXiv}[1]{\href{https://arxiv.org/abs/#1}{\nolinkurl{https://arxiv.org/abs/#1}}}

\bibitem[{Abbasi {et~al.}(2022)Abbasi, Ackermann, Adams, Aguilar, Ahlers, Ahrens, Alameddine, Alispach, Alves~Jr, {et~al.}}]{icecube2022evidence}
Abbasi, R., Ackermann, M., Adams, J., {et~al.} 2022, Science, 378, 538

\bibitem[{Abdollahi {et~al.}(2022)Abdollahi, Acero, Baldini, Ballet, Bastieri, Bellazzini, Berenji, Berretta, Bissaldi, Blandford, Bloom, Bonino, Brill, Britto, Bruel, Burnett, Buson, Cameron, Caputo, Caraveo, Castro, Chaty, Cheung, Chiaro, Cibrario, Ciprini, Coronado-Blázquez, Crnogorcevic, Cutini, D’Ammando, Gaetano, Digel, Lalla, Dirirsa, Venere, Domínguez, Ramazani, Fegan, Ferrara, Fiori, Fleischhack, Franckowiak, Fukazawa, Funk, Fusco, Galanti, Gammaldi, Gargano, Garrappa, Gasparrini, Giacchino, Giglietto, Giordano, Giroletti, Glanzman, Green, Grenier, Grondin, Guillemot, Guiriec, Gustafsson, Harding, Hays, Hewitt, Horan, Hou, Jóhannesson, Karwin, Kayanoki, Kerr, Kuss, Landriu, Larsson, Latronico, Lemoine-Goumard, Li, Liodakis, Longo, Loparco, Lott, Lubrano, Maldera, Malyshev, Manfreda, Martí-Devesa, Mazziotta, Mereu, Meyer, Michelson, Mirabal, Mitthumsiri, Mizuno, Moiseev, Monzani, Morselli, Moskalenko, Negro, Nuss, Omodei, Orienti, Orlando, Paneque, Pei, Perkins, Persic, Pesce-Rollins, Petrosian,
  Pillera, Poon, Porter, Principe, Rainò, Rando, Rani, Razzano, Razzaque, Reimer, Reimer, Reposeur, Sánchez-Conde, Parkinson, Scotton, Serini, Sgrò, Siskind, Smith, Spandre, Spinelli, Sueoka, Suson, Tajima, Tak, Thayer, Thompson, Torres, Troja, Valverde, Wood, \& Zaharijas}]{Abdollahi_2022}
Abdollahi, S., Acero, F., Baldini, L., {et~al.} 2022, ApJS, 260, 53

\bibitem[{{Ackermann} {et~al.}(2012)}]{a5:fermi_starbursts2012}
{Ackermann}, M., {et~al.} 2012, The Astrophysical Journal, 755, 164

\bibitem[{{Beck} {et~al.}(2005)}]{beck2005revised1}
{Beck}, R., {et~al.} 2005, Astronomische Nachrichten: Astronomical Notes, 326, 414

\bibitem[{Cavagnolo {et~al.}(2010)Cavagnolo, McNamara, Nulsen, Carilli, Jones, \& B{\^\i}rzan}]{cavagnolo2010relationship}
Cavagnolo, K., McNamara, B., Nulsen, P., {et~al.} 2010, The Astrophysical Journal, 720, 1066

\bibitem[{{Dermer} \& {Menon}(2009)}]{Dermer+2009_book}
{Dermer}, C.~D., \& {Menon}, G. 2009, {High Energy Radiation from Black Holes: Gamma Rays, Cosmic Rays, and Neutrinos} (Princeton: Princeton University Press)

\bibitem[{Eichmann {et~al.}(2022)Eichmann, Oikonomou, Salvatore, Dettmar, \& Tjus}]{eichmann2022solving}
Eichmann, B., Oikonomou, F., Salvatore, S., Dettmar, R.-J., \& Tjus, J.~B. 2022, The Astrophysical Journal, 939, 43

\bibitem[{{Gallimore} {et~al.}(2004){Gallimore}, {Baum}, \& {O'Dea}}]{Gallimore+2004}
{Gallimore}, J.~F., {Baum}, S.~A., \& {O'Dea}, C.~P. 2004, The Astrophysical Journal, 613, 794

\bibitem[{Garc{\'\i}a-Burillo {et~al.}(2014)Garc{\'\i}a-Burillo, Combes, Usero, Aalto, Krips, Viti, Alonso-Herrero, Hunt, Schinnerer, Baker, {et~al.}}]{garcia2014molecular}
Garc{\'\i}a-Burillo, S., Combes, F., Usero, A., {et~al.} 2014, Astronomy \& Astrophysics, 567, A125

\bibitem[{Ho(2008)}]{Ho2008}
Ho, L.~C. 2008, Annu. Rev. Astron. Astrophys., 46, 475

\bibitem[{{Inoue} {et~al.}(2022){Inoue}, {Cerruti}, {Murase}, \& {Liu}}]{InoueCerrutiMuraseLiu2022}
{Inoue}, S., {Cerruti}, M., {Murase}, K., \& {Liu}, R.-Y. 2022, arXiv e-prints, arXiv:2207.02097

\bibitem[{Inoue {et~al.}(2020)Inoue, Khangulyan, \& Doi}]{Inoue2020}
Inoue, Y., Khangulyan, D., \& Doi, A. 2020, The Astrophysical Journal Letters, 891, L33

\bibitem[{{Kheirandish} {et~al.}(2021){Kheirandish}, {Murase}, \& {Kimura}}]{Kheirandish+2021}
{Kheirandish}, A., {Murase}, K., \& {Kimura}, S.~S. 2021, The Astrophysical Journal, 922, 45

\bibitem[{{Lenain} {et~al.}(2010){Lenain}, {Ricci}, {T{\"u}rler}, {Dorner}, \& {Walter}}]{Lenain+2010}
{Lenain}, J.~P., {Ricci}, C., {T{\"u}rler}, M., {Dorner}, D., \& {Walter}, R. 2010, \aap, 524, A72

\bibitem[{Lopez-Rodriguez {et~al.}(2018)Lopez-Rodriguez, Fuller, Alonso-Herrero, Efstathiou, Ichikawa, Levenson, Packham, Radomski, Almeida, Benford, {et~al.}}]{lopez2018emission}
Lopez-Rodriguez, E., Fuller, L., Alonso-Herrero, A., {et~al.} 2018, The Astrophysical Journal, 859, 99

\bibitem[{Michiyama {et~al.}(2022)Michiyama, Inoue, Doi, \& Khangulyan}]{Michiyama_2022}
Michiyama, T., Inoue, Y., Doi, A., \& Khangulyan, D. 2022, The Astrophysical Journal Letters, 936, L1

\bibitem[{Murase(2022)}]{Murase2022}
Murase, K. 2022, The Astrophysical Journal Letters, 941, L17

\bibitem[{{Murase} {et~al.}(2020){Murase}, {Kimura}, \& {M{\'e}sz{\'a}ros}}]{a5:Murase+2020}
{Murase}, K., {Kimura}, S.~S., \& {M{\'e}sz{\'a}ros}, P. 2020, Phys. Rev. Lett., 125, 011101

\bibitem[{Pacholczyk(1970)}]{pacholczyk1970non}
Pacholczyk, A. 1970, Radio Astrophysics

\bibitem[{Roy {et~al.}(2000)Roy, Wilson, Ulvestad, \& Colbert}]{roy2000slow}
Roy, A.~L., Wilson, A.~S., Ulvestad, J.~S., \& Colbert, E. J.~M. 2000

\bibitem[{{Schlickeiser}(2002)}]{Schlickeiser2002_book}
{Schlickeiser}, R. 2002, {Cosmic Ray Astrophysics} (Berlin: Springer)

\bibitem[{{Seyfert}(1943)}]{Seyfert1943}
{Seyfert}, C.~K. 1943, The Astrophysical Journal, 97, 28

\bibitem[{{Sikora} {et~al.}(1994){Sikora}, {Begelman}, \& {Rees}}]{Sikora+1994}
{Sikora}, M., {Begelman}, M.~C., \& {Rees}, M.~J. 1994, The Astrophysical Journal, 421, 153

\bibitem[{Tully {et~al.}(2009)Tully, Rizzi, Shaya, Courtois, Makarov, \& Jacobs}]{tully2009extragalactic}
Tully, R.~B., Rizzi, L., Shaya, E.~J., {et~al.} 2009, The Astronomical Journal, 138, 323

\bibitem[{Zacharias {et~al.}(2022)Zacharias, Reimer, Boisson, \& Zech}]{Zacharias_2022}
Zacharias, M., Reimer, A., Boisson, C., \& Zech, A. 2022, Monthly Notices of the Royal Astronomical Society, 512, 3948

\end{thebibliography}
\bibliographystyle{aasjournal}

\end{document}